\documentclass[apj]{emulateapj}
\usepackage{natbib}
\usepackage{rotating}
\usepackage{threeparttable}
\usepackage{color}
\usepackage{graphicx}
\usepackage{multirow}

\newcommand{\hi}{H\,{\sc i}}

\shorttitle{The North Polar Spur (Loop I)}
\shortauthors{Sun et al.}

\begin{document}
\title{Faraday Tomography of the North Polar Spur: Constraints on the distance to the Spur and on the Magnetic Field of the Galaxy}

\author{X. H. Sun\altaffilmark{1}, T. L. Landecker\altaffilmark{2}, B. M. Gaensler\altaffilmark{1,3}, E. Carretti\altaffilmark{4,5}, W. Reich\altaffilmark{6}, J. P. Leahy\altaffilmark{7}, N. M. McClure-Griffiths\altaffilmark{8}, R. M. Crocker\altaffilmark{8}, M. Wolleben\altaffilmark{2,9}, M. Haverkorn\altaffilmark{10,11}, K. A. Douglas\altaffilmark{2,12}, A. D. Gray\altaffilmark{2}}
\altaffiltext{1}{Sydney Institute for Astronomy, School of Physics, The University of Sydney, NSW 2006, Australia; X.Sun@physics.usyd.edu.au}
\altaffiltext{2}{National Research Council Canada, Herzberg Program in Astronomy and Astrophysics, Dominion Radio Astrophysical Observatory, P.O. Box 248, Penticton, British Columbia, V2A 6J9, Canada; Tom.Landecker@nrc-cnrc.gc.ca}
\altaffiltext{3}{Dunlap Institute for Astronomy and Astrophysics, The University of Toronto, 50 St. George Street, Toronto, ON M5S 3H4, Canada }
\altaffiltext{4}{CSIRO Astronomy and Space Science, PO Box 76, Epping, New South Wales 1710, Australia}
\altaffiltext{5}{INAF/Osservatorio Astronomico di Cagliari, Via della Scienza 5, I-09047 Selargius, Italy}
\altaffiltext{6}{Max-Planck-Institut f\"{u}r Radioastronomie, Auf dem H\"ugel 69, 53121 Bonn, Germany}
\altaffiltext{7}{Jodrell Bank Centre for Astrophysics, Alan Turing Building, School of Physics and Astronomy, The University of Manchester, Oxford Road, Manchester M13 9PL, UK}
\altaffiltext{8}{Research School of Astronomy and Astrophysics, Australia National University, Cotter Road, Weston Creek, ACT 2611, Australia}
\altaffiltext{9}{Skaha Remote Sensing Ltd, 33 Luxstone Crescent, Airdrie, Alberta, T4B 2W6, Canada}
\altaffiltext{10}{Department of Astrophysics/IMAPP, Radboud University Nijmegen, PO Box 9010, NL-6500 GL Nijmegen, The Netherlands}
\altaffiltext{11}{Leiden Observatory, Leiden University, PO Box 9513, 2300 RA Leiden, The Netherlands}
\altaffiltext{12}{Physics and Astronomy Department, Okanagan College, 1000 KLO Road, Kelowna, British Columbia, V1Y 4X8, Canada}

\begin{abstract} 
We present radio continuum and polarization images of the North Polar Spur 
(NPS) from the Global Magneto-Ionic Medium Survey (GMIMS) conducted with the 
Dominion Radio Astrophysical Observatory 26-m Telescope. We fit polarization 
angle versus wavelength squared over 2048 frequency channels from 1280 to 
1750~MHz to obtain a Faraday Rotation Measure (RM) map of the NPS. Combining 
this RM map with a published Faraday depth map of the entire Galaxy in this 
direction, we derive the Faraday depth introduced by the NPS and the Galactic 
interstellar medium (ISM) in front of and behind the NPS. The Faraday depth 
contributed by the NPS is close to zero, indicating that the NPS is an emitting 
only feature. The Faraday depth caused by the ISM in front of the NPS is 
consistent with zero at $b>50\degr$, implying that this part of the NPS is 
local at a distance of approximately several hundred parsecs. The Faraday depth 
contributed by the ISM behind the NPS gradually increases with Galactic 
latitude up to $b=44\degr$, and decreases at higher Galactic latitudes. This 
implies that either the part of the NPS at $b<44\degr$ is distant or the NPS is 
local but there is a sign change of the large-scale magnetic field. If the NPS 
is local, there is then no evidence for a large-scale anti-symmetry pattern in 
the Faraday depth of the Milky Way. The Faraday depth introduced by the ISM 
behind the NPS at latitudes $b>50\degr$ can be explained by including a 
coherent vertical magnetic field.
\end{abstract}


\section{Introduction}
The North Polar Spur (NPS) is one of the largest coherent structures in the 
radio sky, projecting from the Galactic plane at Galactic longitude 
$l\approx20\degr$ and extending to a very high Galactic latitude 
$b\approx+80\degr$. It was first identified in low frequency radio surveys in 
the 1950s \citep[e.g.][]{hbdh60}. \citet{lqh66} fitted the NPS to part of a 
hypothetical circular structure with a diameter of about $110\degr$ which was 
later named Loop I \citep[e.g.][]{bhs71}.

Observations and theoretical modeling of the NPS up to the 1980s were   
thoroughly reviewed by \citet{sal83}. The NPS had by then been known to have 
(a) strong synchrotron emission whose fractional polarization is very high, up 
to $\sim$70\% at 1.4~GHz, at high latitudes \citep{spo72}, (b) to have strong 
X-ray emission \citep[e.g.][]{bckm72}, (c) to be probably associated with a 
vertical \hi\ filament at $l\sim40\degr$ at velocities around 
0~km~s$^{-1}$ \citep{hj76,cph80}, and (d) to align with starlight polarization 
\citep[e.g.][]{ae76}. All these suggested that the NPS is an old local supernova 
remnant (SNR) at a distance of about 100~pc that has been reheated by the 
shock from a second SNR \citep{sal83}. 

There have been more observations of the NPS at various wavelengths since the  
1980s. From several all-sky radio continuum surveys, the brightness temperature 
spectral index ($T_\nu\propto\nu^\beta$ with $T_\nu$ being the brightness 
temperature at frequency $\nu$) of the NPS is $\beta\approx-2.5$ between 22~MHz 
and 408~MHz~\citep{rcls99} and between 45~MHz and 408~MHz \citep{gmam11}, and 
$\beta\approx-3.1$ between 408~MHz and 1420~MHz \citep{rr88a} at $b>30\degr$ 
where there is little contamination of the diffuse emission from the Galactic 
plane, confirming the NPS as a nonthermal structure.  

The NPS can be clearly seen in the soft X-ray background maps from ROSAT 
observations, particularly in the 0.75~keV band \citep{sef+97}. 
Towards several 
positions, spectra were extracted from observations with ROSAT \citep{ea95},
XMM-Newton \citep{whw+03} and Suzaku \citep{mtb+08} and fit to multiple 
emission components including thermal emission from the NPS. The consensus 
of these papers is that the fraction of the total Galactic \hi\ column density 
in front of the NPS is close to 1 for $b\sim20\degr$ and 0.5 for 
$b\gtrsim30\degr$. Based on the local 3D interstellar medium (ISM) distribution 
from inversion of about 23\,000 stellar light reddening measurements 
\citep{lvv+14} and the corresponding \hi\ column density distribution, 
\citet{plvs14} argued that the NPS is at a distance greater than $\sim$200~pc.

On the other hand, the NPS has also been interpreted as a Galactic scale 
feature. \citet{sof00} proposed that the NPS traces the shock front originating 
from a starburst in the Galactic center about $1.5\times10^7$ years ago. 
\citet{sgc+14} showed that the lower part ($b\leq4\degr$) of the NPS is 
strongly depolarized at 2.3~GHz and thus beyond the polarization horizon of 
about 2--3 kpc. \citet{sof15} found the soft X-ray emission from the lower part 
follows the extinction law caused by the Aquila Rift and derived a lower limit 
of about 1~kpc for the distance to the NPS, although he based this estimate on the 
kinematic distance to the Aquila Rift which has very large uncertainties. Both these 
results suggest that the NPS is a Galactic scale feature. \citet{bc03} 
demonstrated that the NPS can be explained by a bipolar wind from the Galactic 
center. There have also been suggestions \citep[e.g.][]{ktt+13} that the NPS is 
associated with the Fermi Bubble \citep{ssf10}. In contrast, \citet{wol07} 
modelled the NPS as two interacting local shells that can be connected to the 
nearby Sco-Cen association.

A conclusive way to settle the controversy of the nature of the NPS is to 
determine its distance. In this paper we use radio polarization data to 
locate the NPS along the line of sight. We focus on the 1.3--1.8~GHz polarization 
measurements from the Galactic Magneto-Ionic Medium Survey 
\citep[GMIMS;][]{wlh+10}. By comparing the rotation measures (RMs) of the NPS 
emission with those of extragalactic radio sources we establish the contribution 
to Faraday depth by the ISM in front of and behind the NPS, and so constrain its 
distance. The paper is organized as follows. In Sect.~2, we describe the GMIMS 
data and derive the RM map, then scrutinize \hi\ and optical starlight 
polarization data for possible information on the distance to the NPS. In 
Sect.~3, we confine the location of the NPS and discuss the implications for 
modeling of the large scale magnetic field in the Galaxy. We present our 
conclusions in Sect.~4.   

\section{Results}
\subsection{The GMIMS data and the RM map}

GMIMS is a survey of the entire sky with spectro-polarimetry at 
frequencies from 300~MHz to 1.8~GHz using several telescopes in both 
hemispheres \citep{wlh+10}. In this paper we use the data observed 
with the Dominion Radio Astrophysical Observatory 26-m Telescope covering the 
frequency range from 1280 to 1750 MHz split into 2048 channels. Preliminary  
results from GMIMS covering the NPS were shown by \citet{wfl+10}. A 
detailed description of the observations and data processing will be presented 
in a forthcoming paper (Wolleben et al. in prep.). In summary, the observations 
were conducted in long scans along the meridian with a spacing of $12\arcmin$ 
to ensure full Nyquist sampling; a basket-weaving procedure was applied to the 
scans to form all-sky maps at each individual frequency. The data have been 
calibrated to an absolute level. The final data sets are frequency cubes of 
Stokes $I$, $Q$ and $U$ with an angular resolution of $40\arcmin$.  

\begin{figure}
\includegraphics[width=0.45\textwidth]{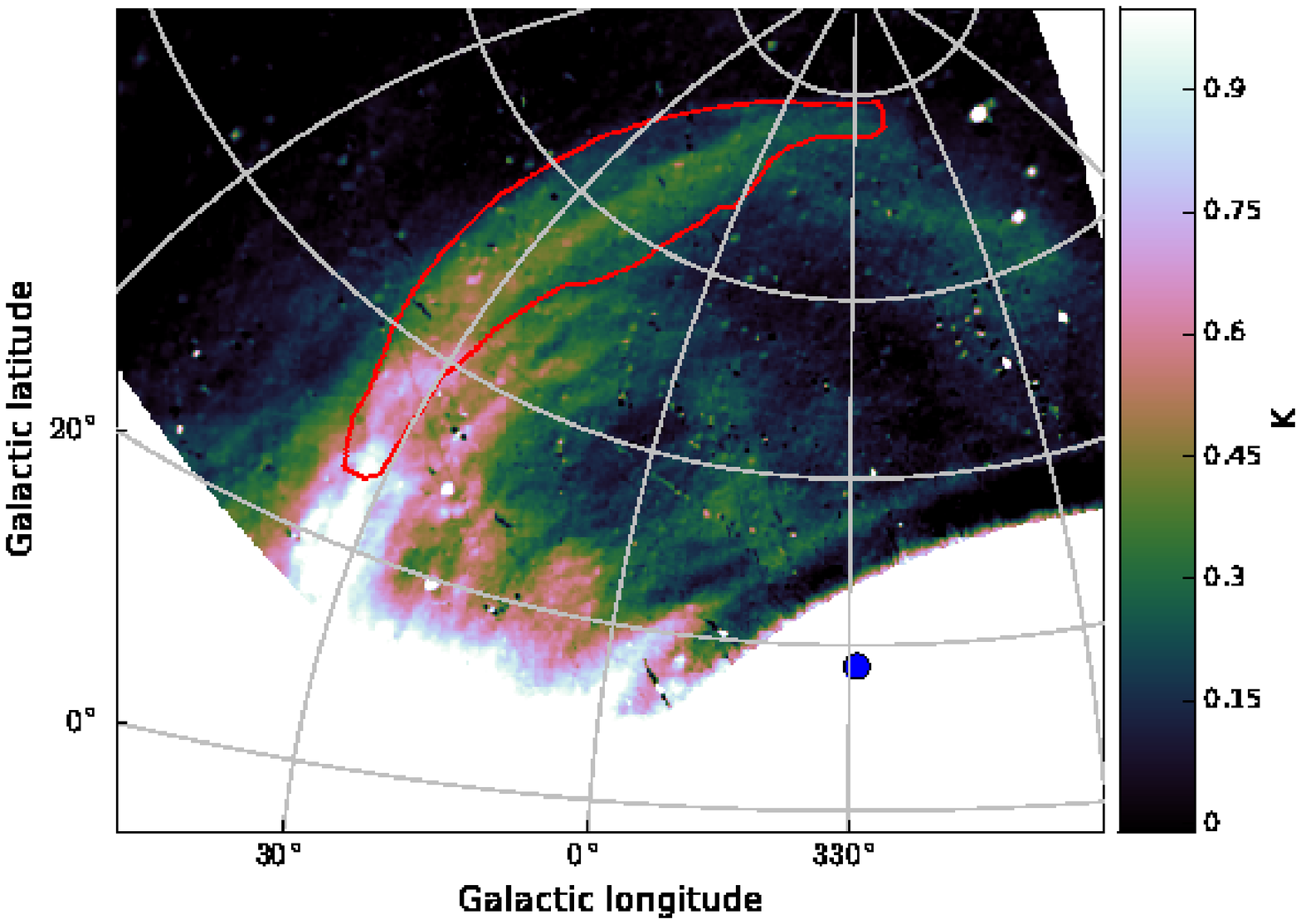}\\[5mm]
\includegraphics[width=0.45\textwidth]{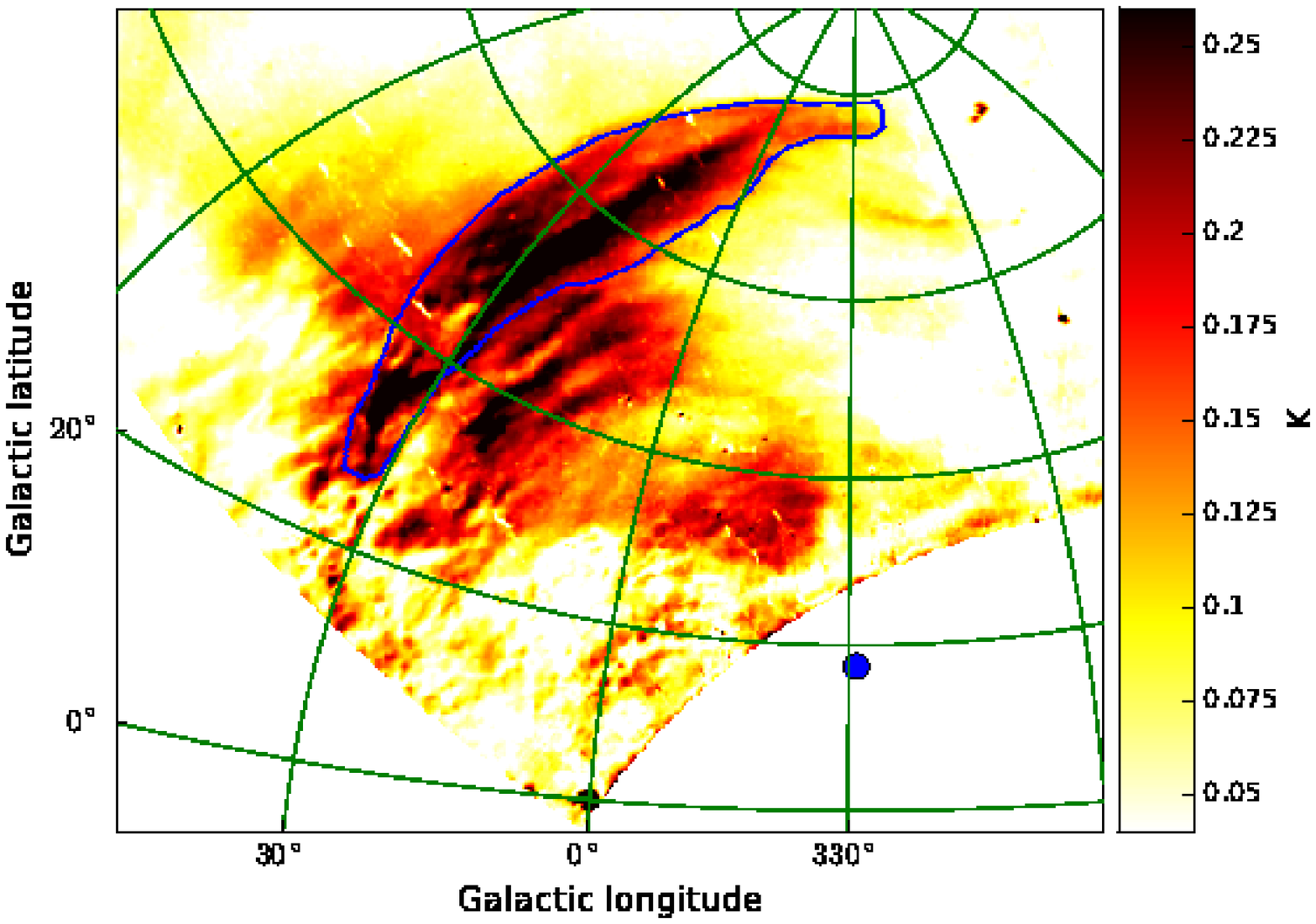}
\caption{Images of the total (upper) and polarized (lower) intensity of the NPS 
averaged between 1.44 and 1.5 GHz. The images are in stereographic projection 
centered at $(l,\,b)=(329\degr,\,17\fdg5)$, the center of Loop I \citep{bhs71}, a position marked as a blue dot in 
each panel. The contour marks the NPS as defined by its polarized intensity. 
The resolution is $40\arcmin$ and the rms noise is about 20~mK for total  
intensity and 6~mK for polarized intensity.}
\label{fig:tppi}
\end{figure}

We selected a frequency range of 1.44 -- 1.5~GHz consisting of 253 channels 
where there is almost no radio frequency interference and derived the average 
total intensity ($I_{1.47}$) and polarized intensity ($P_{1.47}$) over this 
frequency range. The resulting images are shown in Fig.~\ref{fig:tppi} in 
stereographic projection with the projection center at 
$(l,\,b)=(329\degr,\,17\fdg5)$ which is regarded as the center of Loop I 
\citep{bhs71}. We mark a contour denoting the outer boundary of the NPS on 
the basis of its morphology as seen in the $P_{1.47}$ image where  
$P_{1.47}>0.1$~K and RM errors are less than about 5~rad~m$^{-2}$, as 
discussed below. The NPS can be clearly 
identified in both total intensity and polarization. At latitudes higher than 
about $40\degr$, the inner edge of the NPS is much brighter than the outer 
edge, which is consistent with previous observations.

\begin{figure}
\centering
\includegraphics[width=0.45\textwidth]{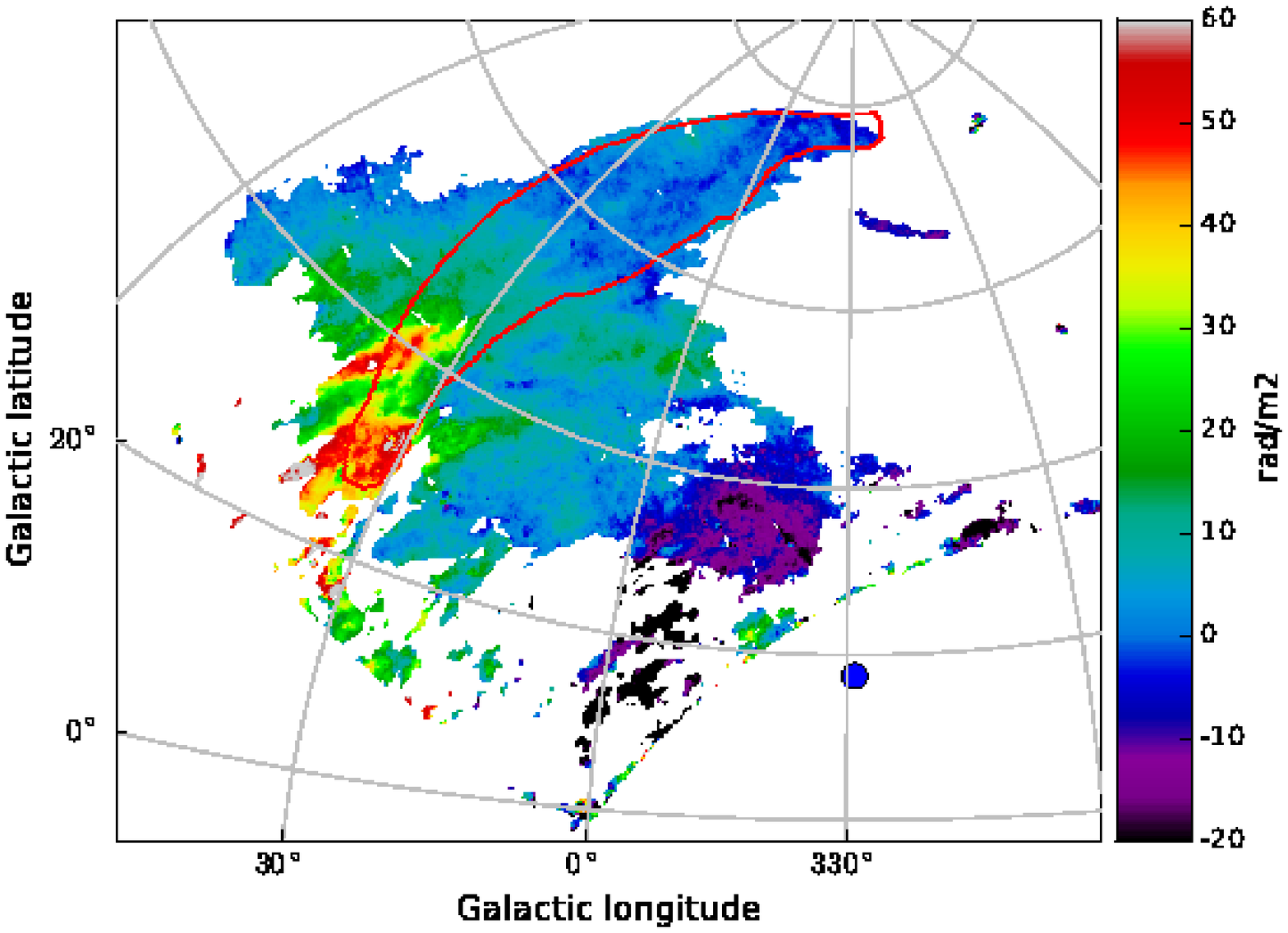}\\[5mm]
\includegraphics[width=0.45\textwidth]{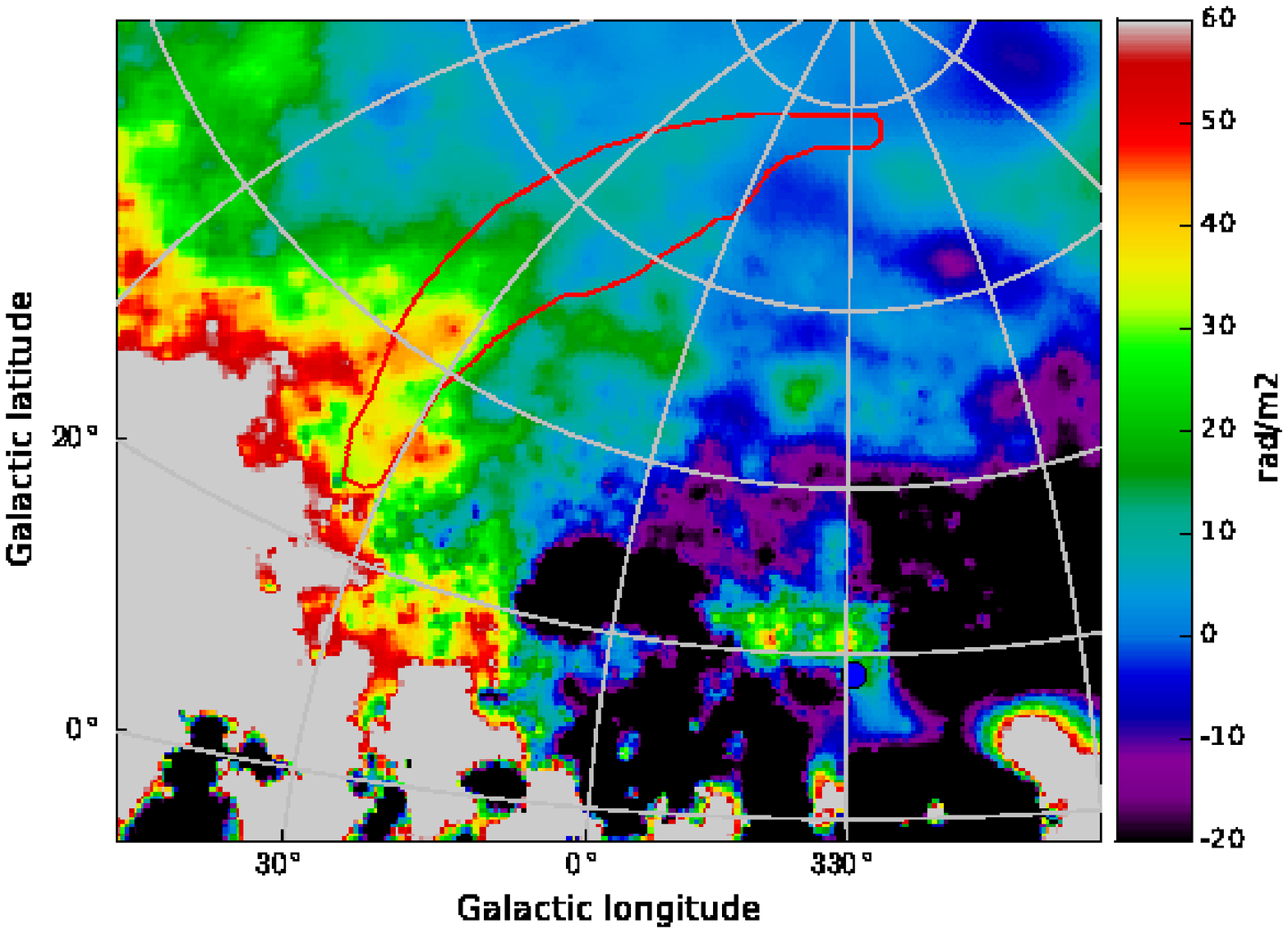}
\caption{RM map for the NPS region from GMIMS (upper panel) and the Galactic 
Faraday depth map constructed by \citet{ojg+15} from RMs of extragalactic 
sources (lower panel). The contours are the same as in Fig.~\ref{fig:tppi}.}
\label{fig:rmmap}
\end{figure}

For each pixel with a polarized intensity $P_{1.47}$ larger than 0.1~K 
(about $5\sigma$-level per frequency channel), we linearly fit polarization 
angles ($\chi$) versus wavelength squared ($\lambda^2$) over the entire 
frequency range from 1280 to 1750~MHz to obtain an RM as
\begin{equation}
\chi(\lambda^2) = \chi_0 + {\rm RM}\lambda^2,
\end{equation}
where $\chi_0$ is a constant. The map of RMs is shown in Fig.~\ref{fig:rmmap} 
(top panel). We also show the Galactic Faraday depth map constructed by 
\citet{ojg+15} from RMs of extragalactic sources in Fig.~\ref{fig:rmmap} (lower 
panel). Although the linear fitting can also be applied to weaker 
polarized intensities, larger errors will be introduced, as shown below. 

It has been demonstrated that the RM from linearly fitting polarization angle 
versus $\lambda^2$ can be wrong unless the behavior of polarization fraction 
against $\lambda^2$ is examined \citep{bur66,frb11}. In reality, the NPS is 
either Faraday thin with only synchrotron-emitting medium or Faraday thick with 
a mixture of synchrotron-emitting and Faraday-rotating medium. For the Faraday 
thin case, the linear relation between polarization angle and $\lambda^2$ 
always holds. For the Faraday thick case, the linear relation holds for 
certain ranges of wavelengths and the RMs represent half of the true values 
\citep[e.g.][]{sbs+98}. For the current data, linear relations between 
polarization angle and $\lambda^2$ can be seen for virtually all the pixels 
with $P_{1.47}$ larger than 0.1~K. The RMs shown in Fig.~\ref{fig:rmmap} (top) 
are thus reliable.

We also generated an RM map using all data over the entire frequency range of 
1280 to 1750~MHz via the RM synthesis method \citep{bd05}. Although the resolution 
in RM is only 150~rad~m$^{-2}$, the signal-to-noise is high, allowing measurements 
of peak RM on finer scales. The resulting RM map is completely  
consistent with the RM map shown in Fig.~\ref{fig:rmmap} (top); the difference 
between RMs calculated in these two ways has a mean value of $0\pm1$~rad~m$^{-2}$ 
over an area much larger than the NPS. We conclude 
that the NPS is Faraday thin as the RM synthesis method often fails to 
reproduce RM when a source is Faraday thick \citep{sra+15}. The RM map in 
Fig.~\ref{fig:rmmap} (top) is very similar to that obtained by \citet{wfl+10} 
from RM synthesis based on the pilot GMIMS data, but has higher resolution and 
sensitivity.
 
\begin{figure}
\centering
\includegraphics[angle=-90,width=0.48\textwidth]{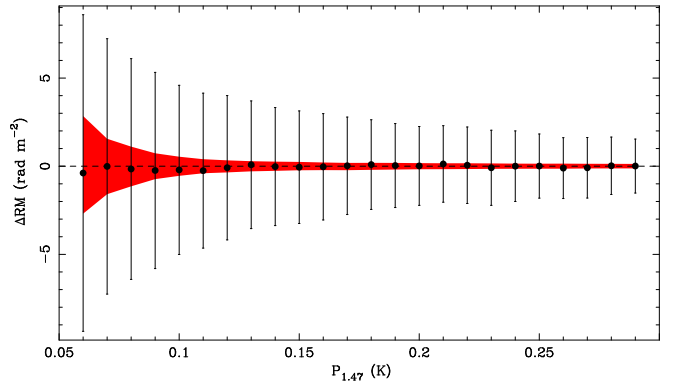}
\caption{Difference between the input and fitted RM values, $\Delta{\rm RM}$, 
versus the polarized intensity averaged over 1.44 -- 1.5~GHz, $P_{1.47}$, from 
faked sources. The red shaded area shows the expected difference at given 
$P_{1.47}$.}
\label{fig:rmtest}
\end{figure}

\begin{figure*}
\centering
\includegraphics[width=0.9\textwidth]{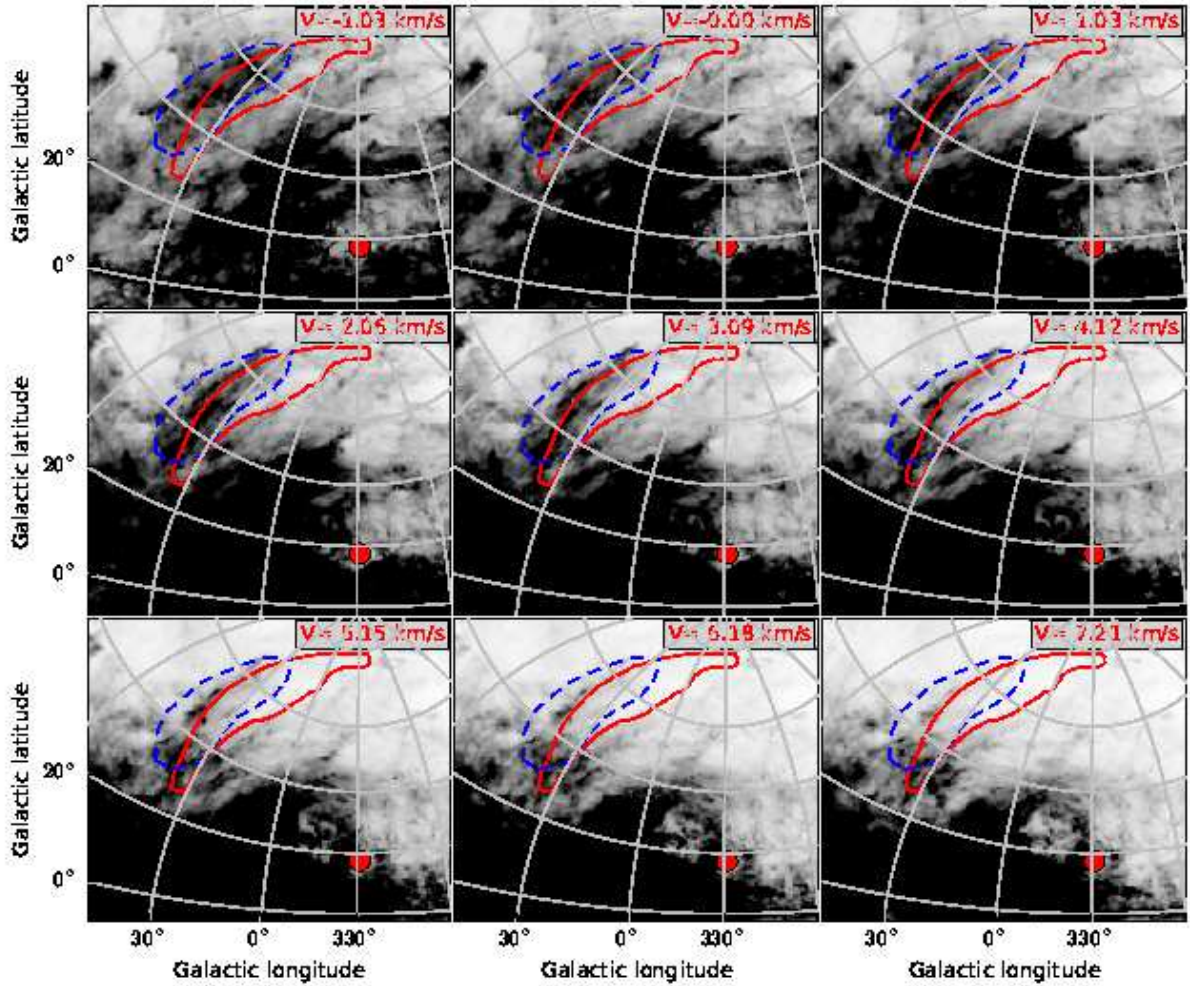}
\caption{\hi\ intensity maps from the LAB survey from velocity 
$-1.03$~km~s$^{-1}$ to $+$7.21~km~s$^{-1}$ with a step of 1.03~km~s$^{-1}$ 
(from top left to bottom right). The gray scale runs from 0.1~K (white) to 
20~K (dark). The NPS is outlined in red solid lines and an \hi\ feature 
possibly associated with the NPS is outlined in blue dashed lines. The red dots 
indicate the projection center as in Fig.~\ref{fig:tppi}.}
\label{fig:himaps}
\end{figure*}

The best published RMs for the NPS are those of \citet{spo84}, based on 
surveys at 408, 465, 610, 820, and 1411~MHz. Polarization angle measurements, 
with different beams at these frequencies, were used to compute RM 
point-by-point; spatial undersampling precluded smoothing to a common 
beamwidth and computation of an RM map. Because of undersampling in frequency, 
$|{\rm{RM}}|$ was restricted to values less than ${35~{\rm{rad~m}}^{-2}}$. The 
resulting RM ``map" is probably sensitive only to spatial scales 
${\gtrsim}{\thinspace}{3^{\circ}}$. No useful comparison of our new RMs 
with these older data is possible.

We made simulations to quantify the RM errors. We extracted a data cube 
centered at an empty area with a size $\sim30\degr\times20\degr$ which contains 
primarily noise but no polarized structures in any of the frequency channels. 
For each pixel, we generated a fake source with a randomly selected intrinsic 
polarization angle, polarized intensity and RM, and added the corresponding $U$ 
and $Q$ of the source to each individual frequency channel. We then derived a 
new data cube of polarization angle and applied the same linear fitting 
procedure as above to calculate a RM map. The difference, $\Delta{\rm RM}$, 
between the derived RM values and the input RM values provides a robust 
estimate of the RM errors. We show $\Delta{\rm RM}$ versus $P_{\rm 1.47}$ in 
Fig.~\ref{fig:rmtest}. 
We repeated the process by adding the same fake sources to Gaussian noise 
with an rms value of 20~mK as measured from the data, 
which yielded the expected errors (red shaded area in 
Fig.~\ref{fig:rmtest}). The real RM errors are much larger than the expected 
errors. This is probably related to low-level scanning effects in the data; the 
basket-weaving process reduces such effects, but does not completely remove 
them.
We kept only those pixels with a $P_{\rm 1.47}$ larger 
than 0.1~K so that the RM errors are lower than about 5~rad~m$^{-2}$.

Two patches with high positive values can be identified in both the RM and 
Faraday depth maps in Fig.~\ref{fig:rmmap}. The one at $b<35\degr$ can be 
clearly seen in the RM map (upper panel in Fig.~\ref{fig:rmmap}), but is not 
especially obvious in the Faraday depth map 
(lower panel in Fig.~\ref{fig:rmmap}). In contrast, the other at 
$36\degr<b<46\degr$ is clearly seen in the Faraday depth map, but has smaller 
extent in the RM map. \citet{wfl+10} attribute both patches to Faraday rotation 
by a local \hi\ bubble associated with Upper 
Scorpius, one of the three subgroups of the Sco-Cen OB association, at a 
distance of about 145~pc. Towards latitudes above about $50\degr$, which are 
not influenced by that \hi\ bubble, RMs gradually decrease to zero with large 
fluctuations. 

\subsection{\hi\ data revisited}\label{sect:hirevisit}

\citet{hj76} claimed that the NPS is associated with \hi\ gas over the velocity 
range $-$20~km~s$^{-1}$ to $+$20~km~s$^{-1}$, using data from the Hat Creek 
\hi\ survey. Using the Leiden/Argentine/Bonn (LAB) survey \citep{kbh+05}, which 
has much higher sensitivity, we re-investigate the associations between \hi\ 
gas and the NPS.       

By comparing the NPS with each individual velocity channel map from the LAB 
survey, we find that a filament oriented almost parallel to $l\approx40\degr$ 
extending from $b\approx30\degr$ to $b\approx70\degr$ over the velocity range 
from $-$1.03 to $+$7.21~km~s$^{-1}$ is possibly morphologically associated with 
the NPS (Fig.~\ref{fig:himaps}), consistent with the finding by \citet{hj76}. 
The vertical \hi\ filament can be best seen at velocity $+2.06$~km~s$^{-1}$, 
roughly parallel to the NPS, gradually fading away towards velocity 
$+7.21$~km~s$^{-1}$ and becoming brighter but mixed with large-scale background 
emission towards velocity $-1.03$~km~s$^{-1}$. A contour based on the 
morphology of the filament at velocity $+2.06$~km~s$^{-1}$ is shown in each 
velocity frame in Fig.~\ref{fig:himaps}.
Because of the high latitude and 
the very low velocity, the distance to the \hi\ structure cannot be constrained.

We estimate the mass of the \hi\ gas 
contained in the region within the dashed blue contour of Fig.~\ref{fig:himaps} 
over the velocity range from $-$1.03 to $+$7.21~km~s$^{-1}$ to be about 
$10^3 D_{100}^2$~M$_\odot$, where $D_{100}$ is the distance to the \hi\ with a 
nominal value of 100 parsecs. Assuming the \hi\ gas is part of a large shell 
structure, \citet{wea79} obtained an expansion velocity of 2~km~s$^{-1}$ which 
corresponds to a kinetic energy of about 
$4\times10^{42}D_{100}^2$ ergs.\footnote{We cannot derive an expansion velocity 
from the data in Fig.~\ref{fig:himaps} because we cannot relate this \hi\ 
feature to other \hi\ filaments to form a large shell structure.} For 
$D_{100}=1$, the kinetic energy is $4\times10^{42}$ ergs, which can be easily 
produced by stellar winds from the Sco-Cen cluster \citep{wea79}, and for 
$D_{100}=100$, the kinetic energy of $4\times10^{46}$ ergs is well below what a 
nuclear explosion \citep{sof00} and galactic winds \citep{bc03} in the Galactic 
center can provide. Thus the \hi\ filament can be either local or far away 
according to the energy budget, and the NPS, if it is associated with the 
\hi\ filament, can be either local or distant. 

\subsection{Optical starlight polarization revisited}\label{sect:starlight}

The light from stars becomes polarized when it is selectively absorbed during 
propagation by dust grains aligned by the magnetic field \citep{dg51}. 
Starlight polarization vectors are parallel to the magnetic field in the dust 
and the polarization fraction depends on the depth of the sightline and on the 
degree of ordering of the magnetic field perpendicular to the sightline 
\citep{flpt02}. In contrast, radio polarization vectors, after correction 
for Faraday rotation, are perpendicular to magnetic field vectors.

\citet{spo72} compared the polarization angles of radio emission at 1415~MHz 
from the NPS with those of optical starlight polarization and found that for 
stars with distances larger than about 100~pc the two angles differ by about 
$90\degr$ indicating that they trace the same magnetic field. This set the 
distance to the NPS at about 100~pc.  

\begin{figure}[!htbp]
\centering
\includegraphics[width=0.48\textwidth]{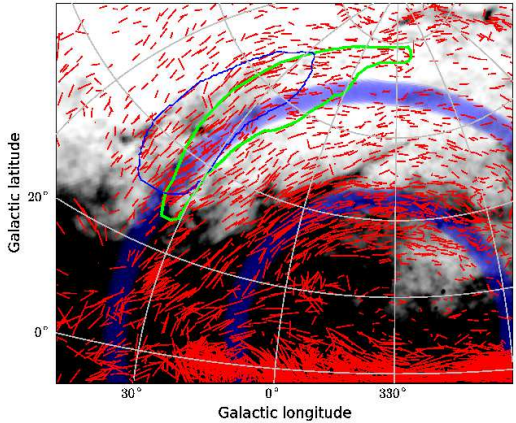}
\caption{Optical starlight polarization vectors (red bars) overlaid on the 
Planck dust image at 353~GHz~\citep{planck_dust14}. The lengths of the bars are 
proportional to the polarization fraction, and their orientations indicate the 
orientations of magnetic fields. The green line and blue line outline the NPS 
and the \hi\ structure described in Sect.~\ref{sect:hirevisit}. The two circles 
are centered at $(l,\,b)=(335\degr,\,10\degr)$ with radii of $35\degr$ and 
$60\degr$.}
\label{fig:star}
\end{figure}

There are now more optical polarization data including the compilations by 
\citet{hei00}, \citet{scr11} and \citet{bpt14}, which motivate us to re-examine 
the correlations between starlight polarization and other tracers of the NPS. 
In Fig.~\ref{fig:star} we show optical starlight polarization data overlaid on 
the Planck dust map~\citep{planck_dust14}.

At latitudes $b>20\degr$ the starlight polarization vectors have a curvature 
which resembles that of the dust structures (Fig.~\ref{fig:star}); the 
curvature of the vectors suggests a center at 
$(l,\,b)\approx(335\degr,\,10\degr)$. In Fig.~\ref{fig:star}, we show two 
partial circles with radii of $35\degr$ and $60\degr$ centered at this 
position. The starlight polarization vectors are in good alignment with the 
circles. There appears to be a dust bubble centered at about the same position 
with a radius of about $30\degr$, but no prominent filamentary structure within this dust bubble.

\citet{bhs71} placed the center of Loop I at 
$(l,\,b)=(329\degr,\,17.5\degr)$, not far from the 
center of starlight polarization vectors. It is thus possible that both the 
NPS and the starlight polarization are products of the same field 
configuration.
The starlight polarization vectors are quite well aligned with the \hi\ feature 
that we identify in Section~\ref{sect:hirevisit}, and, not surprisingly, there 
appears to be dust associated with the \hi\ as well.

We conclude that the starlight polarization vectors cannot be firmly related 
to the NPS on the basis of morphology. We turn now to evidence provided by the 
percentage polarization of the starlight and the relationship between starlight
polarization vectors and radio polarization vectors (which should be orthogonal
if both polarization signals are from the same magnetic field).

\begin{figure}
\centering
\includegraphics[angle=-90,width=0.45\textwidth]{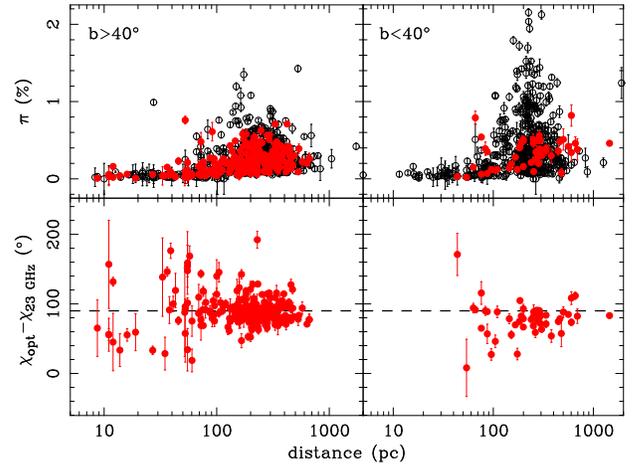}
\caption{Polarization percentage of optical starlight and the difference of 
polarization angles between the optical starlight polarization and WMAP 23~GHz 
polarization, both as a function of distances. The filled circles (red) are for stars towards 
the NPS and the open circles (black) are for stars outside the NPS. The dashed 
lines mark a $90\degr$ polarization angle difference.}
\label{fig:stardist}
\end{figure}

The polarization percentage of the optical starlight polarization versus 
distance to the stars towards and outside the NPS for $b<40\degr$ and $b>40\degr$ is shown 
in Fig.~\ref{fig:stardist}. Most of the distances are from parallax measurements with accuracy better than 50\%. Here ``towards" implies the area within the contour 
denoting the NPS in Figs.~\ref{fig:rmmap}, and ``outside" is 
defined as the area $10\degr$ outside the contour in Fig.~\ref{fig:rmmap}. For directions towards the 
NPS, we also show the polarization angle difference from the WMAP 23~GHz data 
\citep{blw+13} where Faraday rotation is very small.

For $b>40\degr$, the polarization percentages for stars towards and outside the 
NPS are very similar, and they both start increasing at distances above 60~pc, 
reach maximum values between 200 and 300~pc and then slightly decrease up to a 
distance of 700~pc. This can be interpreted by a continuous distribution of 
dust over the distance range 60 -- 700~pc with the magnetic field inside the 
dust gradually changing orientation as a function of distance. The angle 
difference is roughly centered at $90\degr$ for distances larger than about 
60~pc, although the scatter is large. This indicates that the NPS traces 
a similar magnetic field to the dust, and yields a very loose estimate of 
60 -- 700~pc for the distance to the NPS. It is also possible that the NPS is 
further with its magnetic field extending from or coincident with the magnetic field 
in the distance of 60 -- 700~pc.

For $b<40\degr$, there are not many polarization measurements for stars towards 
the NPS. Therefore even a rough estimate of the distance to the NPS is very 
uncertain, and more data are needed. 

\section{Discussion}

\subsection{The location of the NPS}\label{sect:location}

Following \citet{bur66} and \citet{bd05}, we introduce the Faraday depth (FD) 
as a function of distance along a line of sight, $l$, which is defined as
\begin{equation}\label{eq:fd}
{\rm FD}(l) = K\int_{l}^{0} n_e(s) B_\parallel(s) {\rm d}s, 
\end{equation}
where the integral is along the line of sight, $K$ is a constant, $n_e$ is the 
electron density, $B_\parallel$ is the magnetic field projected along the line 
of sight, and $s$ is the distance increment. For the Faraday depth of the 
Galaxy ($\rm FD_G$), $l$ is the distance from the observer to the edge of the 
Galaxy. The differential Faraday depth of a source, $\rm \Delta FD$, can then 
be defined as 
\begin{equation}
{\rm \Delta FD} = {\rm FD}(l_2) - {\rm FD}(l_1), 
\end{equation}
where $l_1$ and $l_2$ are the distance of the near and far boundaries of the  
source, respectively. A detailed discussion of the distinction between RM and 
FD is given by \citet{sra+15}. Throughout the paper, we use the FD map of the 
Galaxy which has been constructed by \citet{ojg+15} primarily based on the RM 
catalog by \citet{tss09} and the RMs towards the Galactic poles by 
\citet{mgh+10}. The extended critical filter \citep{ore11} was used for 
the construction, which was able to simultaneously recover the Faraday depth, 
its angular power spectrum, and the noise co-variance. The minimum scale of the 
final Faraday depth map can be as small as $0.5\degr$. The typical uncertainty is 
about 5~rad~m$^{-2}$ towards latitudes greater than about $40\degr$, and about 
10~rad~m$^{-2}$ towards latitudes between $20\degr$ and $40\degr$.
Note that we also tried the FD map of the Galaxy by 
\citet{xh14} for the analysis below and found similar results. {The FD  
map by \citet{ojg+15} covering the NPS is presented in the lower panel of Fig.~\ref{fig:rmmap}.} 

\begin{figure}
\centering
\includegraphics[angle=-90, width=0.48\textwidth]{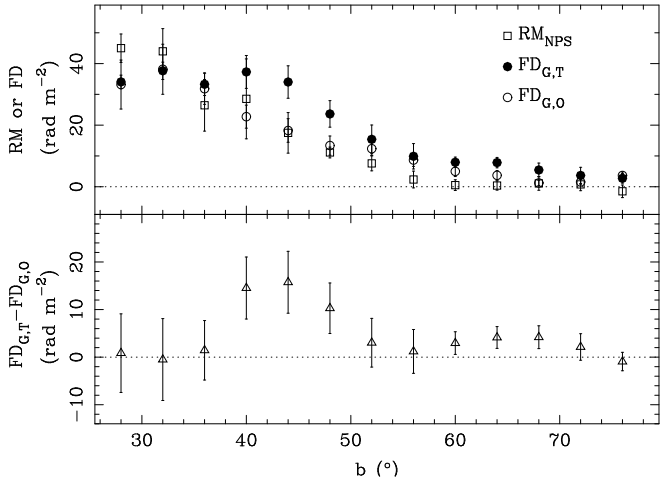}
\caption{Galactic latitude profiles of RM, FD and differential FD. Top panel: 
the filled (open) circles represent FDs of the Galaxy through (outside) the NPS 
and the squares represent RMs of the NPS derived from the GMIMS data; Bottom 
panel: the difference between the FDs through and outside the NPS.}
\label{fig:rmprof}
\end{figure}

Our aim is to constrain the location of the NPS by comparing the differential 
FD of the Galactic ISM in front of the NPS ($\Delta\rm FD_{FG}$) with that of 
the Galactic ISM behind the NPS through to the edge of the Galaxy 
($\Delta\rm FD_{BG}$). The known quantities from observations are the RMs of 
the NPS ($\rm RM_{NPS}$), the FDs of the Galaxy through ($\rm FD_{G,\,T}$) and 
outside ($\rm FD_{G,\,O}$) the NPS; and the unknowns are $\rm \Delta FD_{FG}$, 
$\rm \Delta FD_{BG}$, the differential FD of the NPS itself 
($\rm \Delta FD_{NPS}$) and the differential FD of the \hi\ bubble 
($\rm \Delta FD_{HI}$). The area of the NPS has been outlined in 
Fig.~\ref{fig:rmmap}. We restrict the area outside the NPS to be within 
$10\degr$ longitude from both sides of the NPS at each latitude. 
For Galactic latitudes between $28\degr$ and $76\degr$ we average 
$\rm RM_{NPS}$, $\rm FD_{G,\,T}$, and $\rm FD_{G,\,O}$ in latitude bins of 
$4\degr$ and over all the 
corresponding longitudes and obtain their latitude profiles 
(Fig.~\ref{fig:rmprof}). 

The high positive RMs and FDs that are associated with the local \hi\ bubble in 
front of the NPS \citep[][and their Figure 3]{wfl+10} can be clearly seen in 
Fig.~\ref{fig:rmmap}. Because of the influence of this \hi\ bubble, we divided 
the NPS into two regions:
\begin{itemize}
\item $b\lesssim50\degr$ -- The differential FD of the \hi\ bubble in front of 
the NPS has to be accounted for. We can represent $\rm RM_{NPS}$, 
$\rm FD_{G,\,T}$ and $\rm FD_{G,\,O}$ as
\begin{equation}\label{eq:lower}
\left\{
\begin{array}{rcl}
\rm RM_{NPS} &=& \rm \Delta FD_{HI,\,T} + \Delta FD_{FG} + \frac{1}{2}\Delta FD_{NPS}\\[3mm]
\rm FD_{G,\,T} &=&\rm \Delta FD_{HI,\,T} + \Delta FD_{FG} +  \Delta FD_{NPS} + \Delta FD_{BG} \\[3mm]
\rm FD_{G,\,O} &=&\rm \Delta FD_{HI,\,O} + \Delta FD_{FG} + \Delta FD_{BG}.
\end{array}
\right.
\end{equation} 
Here $\rm \Delta FD_{HI,\,T}$ and $\rm \Delta FD_{HI,\,O}$ are the differential 
FD of the \hi\ bubble covering the NPS and the area outside the NPS, 
respectively. The factor $\frac{1}{2}$ comes from the assumption that the 
thermal gas within the NPS is uniformly mixed with non-thermal emitting gas 
\citep[e.g.][]{sbs+98}. The assumption is reasonable as good linear relations 
between polarization angles and $\lambda^2$ hold towards the NPS. 

\item $b\gtrsim50\degr$ -- There is no influence by the \hi\ bubble in this 
region, and $\rm RM_{NPS}$, $\rm FD_{G,\,T}$ and $\rm FD_{G,\,O}$ can be 
expressed as 
\begin{equation}\label{eq:upper}
\left\{
\begin{array}{rcl}
\rm RM_{NPS} &=& \rm \Delta FD_{FG} + \frac{1}{2}\Delta FD_{NPS}\\[3mm]
\rm FD_{G,\,T} &=&\rm \Delta FD_{FG} + \Delta FD_{NPS} + \Delta FD_{BG}\\[3mm]
\rm FD_{G,\,O} &=&\rm \Delta FD_{FG} + \Delta FD_{BG}.
\end{array}
\right.
\end{equation} 
\end{itemize}

We first estimate the differential FD of the NPS. For the area 
$b\gtrsim50\degr$, it can be derived from Eq.~(\ref{eq:upper}) as 
\begin{equation}\label{eq:fdnps}
\rm \Delta FD_{NPS} = FD_{G,\,T} - FD_{G,\,O}. 
\end{equation}
The results are shown in Fig.~\ref{fig:rmprof} (bottom panel). The average of 
$\rm \Delta FD_{NPS}$ is 2$\pm$4~rad~m$^{-2}$, consistent with zero. This is 
likely due to the lack of thermal electrons as no excess H$\alpha$ emission can 
be seen towards the NPS from the composite all-sky H$\alpha$ map of 
\citet{fin03}. For the area $b\lesssim50\degr$, the differential FD of the NPS 
can be derived from Eq.~(\ref{eq:upper}) as  
\begin{equation}
\rm \Delta FD_{NPS} =  \rm FD_{G,\,T} - FD_{G,\,O} - (\Delta FD_{HI,T}-\Delta FD_{HI,O}).
\end{equation}
The differential FDs of the \hi\ bubble through and outside the NPS are 
unknown, it is therefore difficult to solve for $\rm \Delta FD_{NPS}$. For the 
lower part $b\lesssim36\degr$, $\rm FD_{G,\,T} - FD_{G,\,O}$ is around zero 
(Fig.~\ref{fig:rmprof}, bottom panel), which means
\begin{equation}\label{eqn:fdnpshi}
\rm \Delta FD_{NPS} =   \Delta FD_{HI,O}-\Delta FD_{HI,T}.
\end{equation}
There is no physical connection between the NPS and the \hi\ bubble and hence 
no relation between $\rm \Delta FD_{NPS}$ and $\rm FD_{HI,O}-FD_{HI,T}$. This 
implies that both sides of Eq.~(\ref{eqn:fdnpshi}) are equal to zero for the 
latitude range $b\lesssim36\degr$. Since $\rm \Delta FD_{NPS}$ is close to zero 
towards both larger and smaller Galactic latitudes, we assume it is also close 
to zero towards the middle part $36\lesssim b \lesssim50\degr$. This can be 
corroborated by the fact that the patch with high positive RM crosses the 
eastern edge of the NPS without any change (Fig.~\ref{fig:rmmap}). The large 
values of $\rm FD_{G,\,T} - FD_{G,\,O}$ for $36\lesssim b \lesssim50\degr$ 
(Fig.~\ref{fig:rmprof}) can be attributed to the large difference between 
$\rm \Delta FD_{HI,T}$ and $\rm \Delta FD_{HI,O}$ which can also be seen from 
Fig.~\ref{fig:rmmap}. For the discussions below, we assume 
$\rm \Delta FD_{NPS}$ is zero for $b\lesssim50\degr$.

\begin{figure}
\centering
\includegraphics[angle=-90, width=0.48\textwidth]{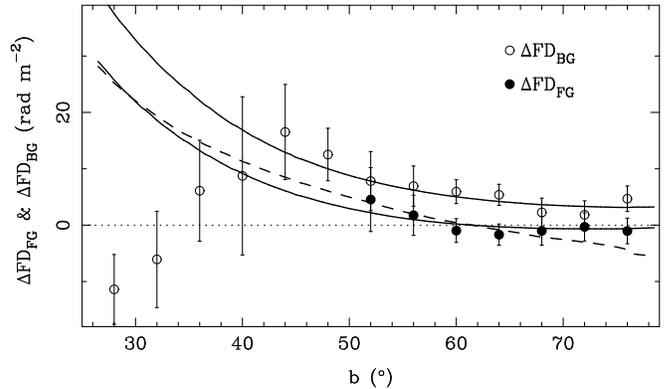}
\caption{$\rm FD_{FG}$ and $\rm FD_{BG}$ versus Galactic latitude. The lower 
solid and dashed lines are from the models by \citet{sr10} and \citet{jf12}, 
respectively. The upper solid line is from the model by \citet{sr10} but with 
an extra dipole field with a strength of 0.2~$\mu$G at the Sun's position.}
\label{fig:fdfgbg}
\end{figure}

We now look at $\rm \Delta FD_{BG}$ and $\rm \Delta FD_{FG}$. For the entire 
latitude range, we can obtain an estimate of $\rm \Delta FD_{BG}$ from 
Eqs.~(\ref{eq:lower}) and (\ref{eq:upper}) as,
\begin{equation}
\rm \Delta FD_{BG} = \rm FD_{G,T}-RM_{NPS}-\frac{1}{2}\Delta FD_{NPS}. 
\end{equation}
From previous discussions, $\rm \Delta FD_{NPS}$ is zero, and 
$\rm \Delta FD_{BG}$ can then be derived. The resulting profile of 
$\rm \Delta FD_{BG}$ is shown in Fig.~\ref{fig:fdfgbg}. We can only solve 
$\rm \Delta FD_{FG}$ for the latitude range $b\gtrsim50\degr$ from 
Eq.~(\ref{eq:upper}) as
\begin{equation}
\rm \Delta FD_{FG} = RM_{NPS} - \frac{1}{2}\Delta FD_{NPS},
\end{equation}
 and the result is shown in Fig.~\ref{fig:fdfgbg}.

For the area $b\gtrsim50\degr$, $\rm \Delta FD_{FG}$ is consistent with 
0~rad~m$^{-2}$. Since $\rm \Delta FD_{BG}$ is not zero, a regular magnetic 
field and constant thermal electron density must exist along the entire line of sight. In this 
case, $\rm \Delta FD_{FG}$ values of zero imply that the path length in the 
integral in Eq.~(\ref{eq:fd}) is close to zero and the NPS is thus a local 
feature. From the 3D modeling of the ISM by \citet{plvs14}, the local cavity, 
defined as an irregularly shaped area of very low density gas surrounding the 
Sun, extends to about 100~pc towards the NPS. From Fig.~\ref{fig:stardist}, 
dust starts to appear only from the distance further than about 60~pc, which 
supports the existence of the local cavity. Within the cavity, the differential 
FD must be around zero, and any contributions to Faraday depths start beyond 
the cavity wall. Since the Galactic magnetic field is predominately parallel to 
the plane \citep[e.g.][and references therein]{srwe08}, the line of sight 
component of magnetic field towards latitudes higher than $50\degr$ is very 
low. The contributions to Faraday depths thus increase very slowly as a 
function of distance. All these considerations place the NPS at a distance of several hundred 
parsecs. 

Towards latitudes $b\lesssim50\degr$, $\rm \Delta FD_{BG}$ increases 
with latitude from a value of $-11$~rad~m$^{-2}$ at $b=28\degr$ to 
17~rad~m$^{-2}$ at $b=44\degr$, and decreases towards higher latitudes.
 There are 
two possible explanations for the behavior of $\rm \Delta FD_{BG}$. One is that 
the NPS is local and $\rm \Delta FD_{BG}$ is from the large-scale Galactic 
magnetic field which has a change of sign at $b=44\degr$. The other is that the 
low latitude part and the high latitude part of the NPS are separate 
structures. It can be seen from the total intensity image in 
Fig.~\ref{fig:tppi} that the low latitude part is much brighter than the high 
latitude part and the transition is not smooth, which can also be seen from the 
recent all-sky radio continuum map at 1.4~GHz by \citet{csb14}. The comparison 
of polarization observations at 2.3 and 4.8~GHz indicates that the very low 
latitude part is further than 2 -- 3~kpc~\citep{sgc+14}. With the low latitude 
part of the NPS far away, the path length from the NPS to the edge of the 
Galaxy is largely reduced and thus $\rm \Delta FD_{BG}$ is much less than that 
extrapolated from high latitudes.

\subsection{Modeling of the Galactic magnetic field}

Modeling of the large-scale magnetic field in the Galaxy is very challenging. 
Ideally models should be able to reproduce a broad range of observations 
such as the FD of the Galaxy, including the total intensity and polarized 
intensity from the synchrotron emission. Both \citet{sr10} and \citet{jf12} 
have built models of the Galactic magnetic field, including both disk and halo 
components.

The differential FD of the Galactic ISM behind the NPS for $b\gtrsim50\degr$ is 
almost equal to the FD of the Galaxy, which allows us to test the models of 
\citet{sr10} and \citet{jf12}. In Fig.~\ref{fig:fdfgbg}, we show the FD profile 
of the Galaxy from both these Galactic magnetic field models. Both models 
predict a FD smaller than $\rm \Delta FD_{BG}$. To increase FDs, the models 
need to have a larger magnetic field along line of sight, which can be achieved 
by increasing either magnetic field parallel to the Galactic plane or magnetic 
field perpendicular to the Galactic plane. From Fig.~\ref{fig:fdfgbg}, it can 
be seen that $\rm \Delta FD_{BG}$ tends to be constant at a value around 
$+$3~rad~m$^{-2}$ for latitudes higher than about $60\degr$, consistent with 
the value obtained by \citet{tss09} from NVSS RMs for area $76\degr<b<90\degr$.
The magnetic field parallel to the Galactic plane cannot contribute FD towards 
the north Galactic pole. Therefore, a vertical component must be incorporated 
to explain the $\rm \Delta FD_{BG}$, which is not included in the model of 
\citet{sr10} and seems insufficient with the X-shape magnetic field in the 
model of \citet{jf12}.     

To demonstrate the necessity of a vertical magnetic field, we tried to add a 
dipole magnetic field component to the model by \citet{sr10}. We find that with
a field strength of 0.2~$\mu$G and a direction pointing from the north Galactic 
pole towards the observer at the position of the Sun, the revised model can now 
reproduce $\rm \Delta FD_{BG}$ for $b\gtrsim50\degr$ (Fig.~\ref{fig:fdfgbg}). 

There is uncertainty in constraining large-scale magnetic field models with 
$\rm FD_{BG}$ at $b\lesssim50\degr$. If the NPS is local, the models by 
\citet{sr10} and \citet{jf12} both fail to reproduce $\rm \Delta FD_{BG}$. 
In this case, the differential FD of the \hi\ bubble dominates the FD of the 
inner Galaxy, producing the mistaken appearance of an anti-symmetric pattern of 
FDs between the first and fourth Galactic quadrants. \citet{sr10} incorporated 
this anti-symmetric pattern into their overall Galactic magnetic field model. 
Subsequently, \citet{wfl+10} highlighted that much of this pattern was due to 
the \hi\ bubble, which led \citet{jf12} to subtract its contribution to 
FD when modeling the Galactic magnetic field. However, there still remain 
high FDs towards the NPS around $b=30\degr$ in the bottom panel of their 
Figure 1 after the subtraction, which is not consistent with our results in 
Fig.~\ref{fig:fdfgbg}. Our work demonstrates that there is no evidence for this 
anti-symmetric pattern in the large-scale FD of the Milky Way at least around 
$b=30\degr$ if the NPS is local.

\section{Conclusions}

The North Polar Spur (NPS), one of the largest coherent structures in the radio 
sky, has been known for more than half a century. The nature of the NPS 
still remains controversial: is it a local supernova remnant or a Galactic 
scale feature related with a starburst or a bipolar wind from the Galactic 
center? We find that it can be both.  

The key to understanding the nature of the NPS is its location in the Galaxy, 
and this has been the focus of our paper. We employed recent \hi\ and 
starlight polarization data and found that neither of these datasets can 
give an exact distance to the NPS, or to the dust structure within the NPS 
perimeter. We then turned to the polarization data from GMIMS for a possible 
constraint of the distance to the NPS. 

GMIMS provides an unprecedented data set with about 2000 frequency channels 
at 1.3 -- 1.8~GHz. Taking advantage of the multi-channel data, we were able to 
obtain an RM map of the NPS by linearly fitting the polarization angle versus 
wavelength squared. Based on the RM map of the NPS and the FD map of the 
Galaxy, we derived the differential Faraday depth of the NPS, the differential 
Faraday depth of the Galactic interstellar medium in front of the NPS and the 
differential Faraday depth of the Galactic interstellar medium behind the NPS through to the edge of the Galaxy for the Galactic latitude range 
$28\degr<b<76\degr$. 

We argue that the part of the NPS at $b\gtrsim50\degr$ is local at a distance 
of about several hundred parsecs because the differential Faraday depth of the 
Galactic ISM in front of the NPS is around zero.  This part of the NPS is 
probably embedded in a large local magnetic field bubble that is traced by 
starlight polarization. With decreasing latitude, differential Faraday depth 
behind the NPS gradually increases, reaches a maximum at $b=44\degr$, and then 
slowly decreases. This implies that either the NPS at $b<44\degr$ is far away 
or the NPS is local but the large-scale magnetic field has a sign change. If the 
NPS is local, the large positive Faraday depths are contributed by an \hi\ bubble 
in front of the NPS, and the large-scale anti-symmetric pattern in Faraday depth 
is then not contributed by a large-scale magnetic field.

We show that the Galactic magnetic field models by \citet{sr10} and 
\citet{jf12} cannot reproduce the differential Faraday depth behind the NPS at 
$b>50\degr$. We find that the model by \citet{sr10} plus a dipole magnetic 
field with a direction pointing from the north to the south Galactic pole and a 
strength of 0.2~$\mu$G at the Sun's position can explain the differential 
Faraday depth behind the NPS. This demonstrates that there exists a coherent 
large-scale vertical magnetic field in the Galaxy near the Sun's position, 
which should be taken into account in future modeling of Galactic magnetic 
fields. 

The location of the NPS is uncertain because the differential Faraday depth in 
front of the NPS cannot be solved at $b\lesssim50\degr$ due to the 
contamination of a local \hi\ bubble in front of the NPS. Future polarimetric 
observations at lower frequencies that provide a higher resolution in Faraday 
depth are needed to properly account for the Faraday depth of the \hi\ bubble. 

\begin{acknowledgements}
The Dominion Radio Astrophysical Observatory is operated as a national facility 
by the National Research Council Canada. We thank Roland Kothes for useful discussions and Patricia Reich for careful reading of the manuscript. XHS and BMG were supported by 
the Australian Research Council through grant FL100100114. MW and KAD were 
supported by the Natural Sciences and Engineering Research Council, Canada.
\end{acknowledgements}

\bibliography{/import/fenway2/xhsun/bibtex.bib}
\end{document}